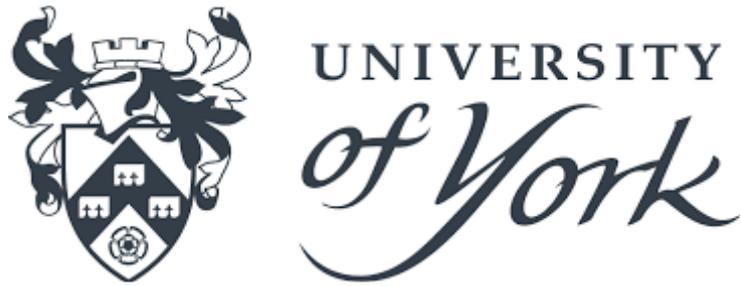

# Time-limited Metaheuristics for Cardinality-constrained Portfolio Optimisation

Alexander Nikiporenko

Supervisor: Oluwaseyi Oginni

January 2023

Independent Research Project submitted in part fulfilment of the degree of MSc in Computer Science with Artificial Intelligence

8798 words

Department of Computer Science

University of York

# Acknowledgements

I would like to thank the following people for helping with this research project:

My supervisor, Oluwaseyi Oginni, for providing guidance and feedback throughout the project.

Saul Cross, and others involved in creating and running the Applied Artificial Intelligence module, which inspired me to explore metaheuristics.

Waseem Ahmad for his guidance during the creation of the research proposal.

My partner, Anicée Landao, for her patience and encouragement.

# Executive Summary

A financial portfolio contains assets that offer a return with a certain level of risk. To maximise returns or minimise risk, the portfolio must be optimised – the ideal combination of optimal quantities of assets must be found. The number of possible combinations is vast. Furthermore, to make the problem realistic, constraints can be imposed on the number of assets held in the portfolio and the maximum proportion of the portfolio that can be allocated to an asset. This problem is unsolvable using quadratic programming, which means that the optimal solution cannot be calculated.

A group of algorithms, called metaheuristics, can find near-optimal solutions in a practical computing time. These algorithms have been successfully used in constrained portfolio optimisation. However, in past studies the computation time of metaheuristics is not limited, which means that the results differ in both quality and computation time, and cannot be easily compared.

This study proposes a different way of testing metaheuristics, limiting their computation time to a certain duration, yielding results that differ only in quality. Given that in some use cases the priority is the quality of the solution and in others the speed, time limits of 1, 5 and 25 seconds were tested. Three metaheuristics – simulated annealing, tabu search, and genetic algorithm – were evaluated on five sets of historical market data with different numbers of assets.

Although the metaheuristics could not find a competitive solution in 1 second, simulated annealing found a near-optimal solution in 5 seconds in all but one dataset. The lowest quality solutions were obtained by genetic algorithm.

# Table of Contents

# List of Tables

# List of Figures

# 1 Introduction

In finance, a portfolio is a group of assets that are acquired for the purpose of making a profit. Modern portfolio theory, developed by Markowitz[1][2], addresses portfolio optimisation by comparing the mean return of a portfolio with the variance of its returns. The goal is to allocate the assets in such a way that the returns are maximised, and the risk is minimised. Given that risk tolerance varies from one investor to another, there is a set of optimal portfolios that offer the maximum expected return for the required level of risk, known as the efficient frontier[3].

The classical Markowitz model is unconstrained, where a portfolio can contain an unlimited number of assets that can be traded in any quantity[1]. To make the problem more realistic, constraints can be added that specify the number of assets in the portfolio and limit the proportion of the portfolio that a single asset can occupy. However, these constraints change the problem from one that can be solved by quadratic programming, to an nondeterministic polynomial time-hard (NP-hard) quadratic mixed-integer problem[4]. Solving such a problem using exact methods would require extremely long computation times, making this approach impractical in the context of stock trading, where decisions are made quickly[5]. In fact, the introduction of Hybrid Market by the New York Stock Exchange in 2006 resulted in order execution times being reduced from 10 seconds to less than one second[6].

An alternative is offered by metaheuristics, which are domain-independent optimisation algorithms that generate near-optimal solutions in a practical computation time[7]. The literature identifies two categories of metaheuristics: population-based metaheuristics (P-metaheuristics), which maintain multiple candidate solutions, and single-solution metaheuristics (S-metaheuristics), which work on a single solution[8].

S-metaheuristics converge faster than P-metaheuristics due to their lower computational complexity, but they are also more likely to become stuck in a local optimum[9]. Recent surveys point out that S-metaheuristics represent a gap in research, as P-metaheuristics have been the focus of more studies[7][10]. Therefore, this study implements two S-metaheuristics, simulated annealing (SA) and tabu search (TS); and one P-metaheuristic, genetic algorithm (GA), which is used as a baseline.

Although the application of metaheuristics to the portfolio optimisation problem is a common research topic, most studies terminate the metaheuristics after a specific number of iterations[11]–[14]. Given that each implementation of a metaheuristic has a different duration per iteration, this approach leads to different values for both solution quality and time, making it challenging to compare their performance. For example, there is no clear answer to the question of whether an algorithm that achieves a better solution, but takes longer to execute, provides better performance. Therefore, this paper proposes a different approach to testing the metaheuristics, terminating the algorithms after a specific period of time. As a result, metaheuristics can be evaluated solely on the quality of their solutions, as their execution time is the same.

The last decade has seen a marked acceleration in the pace of trading in financial markets[15], so the time it takes for metaheuristics to reach a solution is crucial. This study therefore focuses on the short-term performance of metaheuristics, using time limits of 1, 5 and 25 seconds. Identifying which metaheuristics perform best under which time limits is beneficial in real-world applications – in some use cases, the quality of the solution is key, and in others, speed. Although the authors are not aware of other studies of time-limited metaheuristics in the context of portfolio optimisation, S-metaheuristics have outperformed P-metaheuristics in other time-limited contexts[16].

The research question can therefore be formulated as follows: in time-limited, constrained, portfolio optimisation problem, can S-metaheuristics produce higher solution quality than P-metaheuristics? Since this study uses the performance of P-metaheuristics as the baseline, the null hypothesis is that S-metaheuristics provide lower quality solutions than P-metaheuristics for the same computation time.

The aim of this study is to investigate the impact of time limits (1, 5, 25 seconds) on the performance of metaheuristics (SA, TS, GA) in the context of constrained portfolio optimisation. The objectives are: implement metaheuristics in Python; define and tune hyperparameters to optimise algorithm performance; test metaheuristics on historical market data.

## 2   Background Work

### 2.1  Metaheuristics

Given that solving real-world optimisation problems using exact methods requires impractical time, approximate algorithms have been developed as a practical alternative. There are two categories of approximate algorithms: heuristics and metaheuristics[17]. Whilst heuristics are problem-dependent, metaheuristics are domain-independent, which means that they can be adapted to solve any optimisation problem. Metaheuristics are designed to solve large problems quickly by effectively exploring the search space[17].

Metaheuristics can be categorised into S-metaheuristics and P-metaheuristics[18]. S-metaheuristics focus on exploitation or intensification by iteratively searching the neighbourhood of a single solution[10]. SA and TS, S-metaheuristics implemented in this study, are based on an algorithm called hill climbing – a greedy algorithm that iteratively selects the best neighbouring solution. The key weakness of hill climbing is that it can get stuck at the local optimum[19]. The main features of SA and TS, discussed in their respective chapters, are designed specifically to overcome this weakness.

P-metaheuristics focus on exploration or diversification by exploring a relatively large search space consisting of multiple solutions that form a population[10]. The most studied P-metaheuristics are evolutionary algorithms (such as GA) inspired by Darwin's theory of evolution[17]. The algorithms simulate the evolution of a population, with reproduction and survival of the fittest being key concepts[20]. However, there is a cost to maintaining a population of solutions, which is the $O(MN^3)$ computational complexity[21].

An in-depth discussion of metaheuristics can be found in [17], a comprehensive textbook which includes pseudocode for most of the known metaheuristics. Flowcharts of the key algorithms can be found in [22].

## 2.2 Portfolio Optimisation

The use of metaheuristics for portfolio optimisation has been the subject of several surveys, which provide an overview of general trends in this area. According to [10], the most popular S-metaheuristics are SA and TS, while the most popular P-metaheuristics are GA and particle swarm optimisation (PSO). Although P-metaheuristics are more frequently used, the survey claims that their superior performance has not yet been demonstrated[10]. The need for quicker and simpler metaheuristics is also mentioned, and two S-metaheuristics – iterated local search (ILS) and variable neighbourhood search (VNS) – are suggested as potential research avenues[10].

The authors do not discuss time complexity in their summaries of the surveyed papers; instead, they focus on the comparative quality of the solutions. The fact that the execution time of the metaheuristics is crucial in real-world scenarios[15] points to an issue with this approach – if the algorithm takes too long to find a good solution, it will lose relevance. In fact, another survey[7] notes that the articles reviewed often lack information on the computational power used and execution time required, even though computational complexity is a crucial aspect. This suggests that there is a research gap in understanding how to balance the quality and speed of metaheuristics.

According to [7], S-metaheuristics are underutilised and their parameters are simpler to configure than those of P-metaheuristics. Since it is usually easier to understand the internal behaviour of S-metaheuristics, the authors also comment on the transparency of S- and P-metaheuristics, calling them 'white box' and 'black box' methods respectively. Due to material risk, the concept of transparency is important for financial applications of metaheuristics – the user needs to understand what caused the portfolio to lose value[23].

Ant colony optimisation (ACO) is another frequently used P-metaheuristic[7]. Its popularity has given rise to other P-metaheuristics involving various social insect species, which [24] characterises as puzzling. [24] argues that metaphors drawn from natural phenomena may lead to confusing terminology that masks similarities with more traditional approaches. In addition, [24] draws attention to the fact that some of these novel metaheuristics only perform well by excessively fine-tuning the hyperparameters, which leads to poor generalisation to real-world contexts. Therefore, it must be ensured that metaheuristics are generalisable and transparent.

Another survey[25] concludes that the current state of research does not identify the best metaheuristic for portfolio optimisation. Key factors in the development of a successful metaheuristic approach are identified: careful choice of the objective function; selection of an appropriate programming language; and systematic evaluation of different design options[25].

A detailed analysis of the problems encountered in real-world financial applications is presented in [3], a book on portfolio optimisation which briefly discusses metaheuristics. [26] is a thorough guide to the various computational techniques used in finance. Of particular interest is a discussion of backtesting – modelling an investment strategy using past data. Although backtesting is an excellent tool for evaluating an investment approach, it can lead to overfitting[26].

## 2.3 Related Work

Numerous studies compare the performance of metaheuristics in the context of constrained portfolio optimisation. [12] is the first paper of this kind – an influential work, which datasets have since been used in numerous other papers[13][27]–[30], including this study. In [12], GA, TS, and SA are applied to both constrained and unconstrained optimisation. [Tab.1] shows the results of constrained optimisation for all datasets, while [Tab.2] shows the results of unconstrained optimisation for the Nikkei dataset only. Although absolute time measurements are no longer applicable due to technological advances since the publication of the study, comparative times are still relevant.

Tab.1: Cardinality-constrained results from [12].

| Index | Number of assets | | SA | TS | GA |
|---|---|---|---|---|---|
| Hang-Seng | 31 | MPE | 1.0957 | 1.1217 | 1.0974 |
|  |  | Time (s) | 79 | 74 | 172 |
| DAX-100 | 85 | MPE | 2.9297 | 3.3049 | 2.5424 |
|  |  | Time (s) | 210 | 199 | 544 |
| FTSE-100 | 89 | MPE | 1.4623 | 1.6080 | 1.1076 |
|  |  | Time (s) | 215 | 246 | 573 |
| S&P-100 | 98 | MPE | 3.0696 | 3.3092 | 1.9328 |
|  |  | Time (s) | 242 | 225 | 638 |
| Nikkei-225 | 225 | MPE | 0.6732 | 0.8975 | 0.7961 |
|  |  | Time (s) | 553 | 545 | 1964 |
| Average |  | MPE | 1.8461 | 2.0483 | 1.4953 |

Tab.2: Unconstrained results from [12].

| Index | Number of assets | | SA | TS | GA |
|---|---|---|---|---|---|
| Nikkei-225 | 225 | MPE | 1.7681 | 15.9163 | 0.0085 |
|  |  | Time (s) | 281,588 | 210,929 | 227,220 |

Comparing the constrained results for the Nikkei dataset, it can be observed that although mean percentage errors (MPEs) are broadly similar for the three metaheuristics used, the execution time of GA, the only implemented P-metaheuristic, is almost four times as long as that of S-metaheuristics. Meanwhile, the results for the unconstrained problem show that while the execution time is similar for the three metaheuristics, GA's MPE is significantly smaller. Therefore, a conclusion on the trade-off between the quality and speed of metaheuristics cannot be made on the basis of this study.

The same datasets and metaheuristics are investigated in [13]. However, here metaheuristics contain an additional subset optimisation step based on a quadratic equation. Given that the studies use different hardware, the time measurements in [Tab.3] have been multiplied by 70 to make them comparable to [Tab.1], as recommended in [31]. The results[Tab.3] are markedly different: the performance of TS and SA has declined significantly, while GA outperforms these two S-metaheuristics in both speed and quality. Therefore, it can be noted that while GA offers superior performance in this study, the implementation of the algorithms has a significant impact on their performance.

Tab.3: Cardinality-constrained results from [13].

| Index | Number of assets | | SA | TS | GA |
|---|---|---|---|---|---|
| Nikkei-225 | 225 | MPE | 1.1193 | 0.7838 | 0.3353 |
| | | Time (s) | 42,280 | 28,980 | 7,280 |

In [32], SA, TS, GA and a two-phase local search (2PLS) solve constrained portfolio optimisation problem, using both real and artificial datasets. Two real datasets contain 40 and 190 assets respectively, and are therefore similar to the datasets used in this study. The artificial datasets are significantly larger, containing 1,416 assets. Instead of finding the efficient frontier of the portfolios, as is done in the two above-mentioned papers and in this study, a single portfolio is selected for each dataset using utility functions that calculate different financial indicators[32]. GA achieves the best results, while SA and 2PLS achieve competitive results only in the real dataset. TS achieves the worst results in all cases. The algorithms are not time-limited, resulting in significant differences in computation time, from 2 seconds (SA) to 11 seconds (TS) on the artificial dataset[32].

Although [16] applies metaheuristics to the travelling salesman problem, it is relevant to this research because the algorithms terminate after a certain period of time, which is also the case in this paper. This approach differs from the aforementioned

studies[12][13][32], in which metaheuristics are terminated after a certain number of iterations. In [16], six metaheuristics – SA, TS, GA, PSO, harmony search and quantum annealing – are tested different graphs for 100 seconds each. The study finds that S-metaheuristics, particularly SA and TS, outperform P-metaheuristics in a time-limited context[16]. Although these findings relate to a different problem, they illustrate the potential of S-metaheuristics.

## 3   Methodology

This study examines variance and mean return – the summary statistics of solution portfolios, and the time required to run metaheuristics. As these data are numerical, the overall methodology of this paper is quantitative. Given that this study proposes research questions and tests hypotheses with data collected, the epistemology of this study is positivist[33].

The aim of this research is to determine the effect metaheuristics have on the solution quality, which is an experimental approach. The summary statistics of the portfolios are dependent variables in this experiment, while metaheuristics and time limits are independent variables that influence them[34]. The experiment is conducted under laboratory conditions using identical computer hardware and software for all algorithms to ensure that additional factors that may influence the results are controlled[35].

All metaheuristics can be implemented in different ways, which has a significant impact on their performance. The parameters of a particular implementation determine its effectiveness, and so tuning them is an important part of this study. There are two ways to do this: by using the recommended standard values described in past studies, or by trial and error[22]. In this paper, the approaches are combined: the initial standard values are adjusted by trial and error.

The performance of metaheuristics can be evaluated using real-time market data, which is impractical for multiple reasons. Given that this study is conducted on a single computer and that stock market prices are typically updated every minute[36], real-time market data would lead to each metaheuristic being tested on a different dataset, making it impossible to adequately compare results. One way around this problem is to simultaneously test the metaheuristics on a number of identical computers, which would significantly increase the cost of this study.

An alternative approach chosen for this study is to evaluate the metaheuristics using backtesting, defined as the use of past data to model an investment strategy[26]. Backtesting involves a number of risks, the most significant of which is overfitting, which results in the algorithms only performing well on a specific dataset[37]. To prevent this, the metaheuristics are tested on five datasets containing different numbers of assets. Furthermore, given that all the metaheuristics chosen for this study make stochastic decisions, the quality of their solutions is not consistent[22]. Therefore, each algorithm is tested several times and the results are averaged.

## 4 Experimental Setup

### 4.1 Datasets

Metaheuristics are tested on five datasets with different numbers of stocks. Each dataset is collected from a particular stock market index, representing a segment of the financial market[38]. The five selected datasets correspond to the following indices (number of stocks in brackets): Hang-Seng (31), DAX-100 (85), FTSE-100 (89), S&P-100 (98) and Nikkei-225 (225). The indices were recorded weekly in 1992-1997 and stocks with missing values were removed[28].

These datasets, which are a part of the publicly accessible OR-Library[39] and are available for free under the permissive MIT licence[40], were produced for [12]. As such, they are classified as secondary data. The use of secondary data provides significant savings in time and labour compared to the collection of first-hand data[41]. Furthermore, these datasets have been used in other studies dealing with portfolio optimisation[13][27]–[30]. The availability of these studies makes it possible to verify the validity of the results by comparing them with past studies.

In line with modern portfolio theory[1], solution portfolios are evaluated by their mean return and measures of risk (variance and standard deviation). As these statistics are collected specifically for the purpose of this study, they are classified as primary data[42].

OR-Library also provides the values that make up the unconstrained efficient frontier (UEF) for each of the datasets[43]. The frontiers are not continuous, but consist of 2,000 pairs of values representing the mean return and the corresponding variance. These values have been calculated using Numerical Algorithms Group library routine E04NAF[44] for evaluating the quality of solution portfolios by measuring their deviation from the UEF[12]. Several other studies use the same approach to evaluate metaheuristics[13][30]. To ensure that the results of this paper are comparable with those of past studies, the performance of metaheuristics is evaluated in the same way[45].

## 4.2 Objective Function

The aim of optimisation problems is to find a solution that minimises or maximises the objective function[22]. Careful choice of the objective function is key to the success of a metaheuristic[25]. Given that in this paper solution portfolios are evaluated by measuring their deviation from the UEF, one possible objective function could be the Euclidean distance of the portfolio from the UEF. However, the disadvantage of this approach is that it is only applicable if a pre-calculated UEF is available, and therefore it cannot be generalised to the real-world context. Although using a dataset with a known UEF allows the evaluation of implemented methods and comparison with past research, a pre-calculated UEF cannot be expected to exist in a real-life optimisation problem.

Therefore, this study takes another approach, using the objective function[Eq.1] introduced in [12]. This approach does not require a UEF, allowing it to be used in real-world situations. Here, the objective is to minimise a function with a weighting parameter $\lambda$ that specifies the required trade-off between return and risk. The minimum value of this weighting parameter is $\lambda = 0$, at which the function will find the portfolio with the highest return, regardless of the risk involved. At the maximum value of $\lambda = 1$, the function will find the portfolio with the lowest level of risk, regardless of the return. Finding multiple solutions using this objective function with a value of $\lambda$ varying from 0 to 1 will trace the UEF curve[12]. In this study, every metaheuristic is used to find solution portfolios for 50 different values of $\lambda$, uniformly distributed between 0 and 1. This is a common approach used in other studies[12][13][46].

## 4.3 Formulation

The formulation of the cardinality-constrained portfolio optimisation problem used in this paper is a modified version of the problem formulated in [12], which itself derives from the problem in [1].

Let:

$N$ be the number of assets available. This depends on the dataset used.

$K$ be the required number of assets in the portfolio. $K = 10$ in this study.

$\lambda$ be the weighting parameter. 50 values of $\lambda$, uniformly distributed between 0 and 1 in this study.

$\mu_i$ be the expected return of asset $i$.

$\sigma_i$ be the covariance between assets $i$ and $j$.

$w_i$ be the proportion held of asset $i$.

$l$ be the minimum proportion of the portfolio held in asset $i$. $l = 0.01$ in this study.

$\delta_i$ be the zero-one decision variable, which is 1 if asset $i$ is in the portfolio, 0 otherwise.

Minimise
$$\lambda \left[ \sum_{i=1}^{N} \sum_{j=1}^{N} w_i w_j \sigma_{ij} \right] - (1 - \lambda) \left[ \sum_{i=1}^{N} w_i \mu_i \right]$$
Eq.1

Subject to
$$\sum_{i=1}^{N} w_i \delta_i = 1$$
Eq.2

$$\sum_{i=1}^{N} \delta_i = K$$
Eq.3

$$w_i \geq \delta_i l$$
Eq.4

[Eq.1] is the objective function. [Eq.2] requires that the weights of the assets present in the portfolio sum to 1. [Eq.3] specifies the required number of assets in the portfolio. [Eq.4] ensures that each asset present in the portfolio occupies a proportion greater than the required minimum.

## 4.4 Neighbourhood

Given that the search space can be large for complex optimisation problems, one approach is to start locally, by exploring solutions that are measurably close to the initial solution. This set of solutions is called a neighbourhood[19]. Adequate definition of the neighbourhood is the key to the effectiveness of S-metaheuristics[17].

In this paper, the neighbourhood is defined in two dimensions, and the implemented S-metaheuristics alternate moves between the dimensions. Since this study deals with a cardinality-constrained optimisation problem, the constraints help to define the neighbourhood. Given that the solution portfolio must contain $K$ assets[Eq.3], a portfolio cannot be modified by adding a new asset without first removing an asset. Therefore, replacing an asset in the portfolio with an asset that is not in the portfolio is one dimension in which the neighbourhood is defined.

The neighbourhood can also be defined in another dimension by changing the weights of assets already in the portfolio. This operation is subject to two constraints: one specifying the minimum weight of each asset[Eq.4] and the other requiring the sum of all weights in the portfolio to be equal to 1[Eq.2]. One way to achieve this is to select one asset, known as a pivot, and change its weight[47]. As this will also change the sum of all weights, the other weights will need to be adjusted to satisfy the constraint [Eq.2]. Given that [17] recommends defining neighbourhoods by making small changes to the solution, and this approach changes the weight of every asset in the portfolio, it was not chosen for this study as it requires too much change. Instead, the approach is to take two assets in the portfolio and change their respective weights, keeping the sum of their allocations unchanged and ensuring that each weight is greater than the required minimum.

The solutions found in both neighbourhood definitions are all feasible – each found solution satisfies the problem constraints. [25] identifies two other approaches: the repair approach, where an infeasible solution can be found but is then modified to satisfy the constraints[48]; and the penalty approach, where infeasible solutions are accepted with a penalty added to the objective function[49].

## 4.5 Random Portfolios

All metaheuristics implemented in this study feature a step where $n$ random portfolios are generated[C.1]. 1,000 was chosen as the appropriate value for $n$, as it provides balance between speed and chance of generating a portfolio with a sufficiently small objective function. In S-metaheuristics, the initial solution for each value of $\lambda$ is the random portfolio with the smallest objective function for that $\lambda$. This is the approach used in [12]. In [17], a distinction is made between random and greedy approaches to generating an initial solution, noting that random approach is faster but results in a lower quality solution. The approach used in this study can be classified as a hybrid approach, as it selects a locally optimal solution from a pool of random solutions. A similar approach is taken in the implementation of GA – the initial population for each value of $\lambda$ is a specific number of random portfolios with the smallest objective function for the given $\lambda$.

Creating a random portfolio involves two tasks: selecting $K$ random assets from a stock index and allocating weights to the selected assets. While asset selection is a task that can easily be solved using built-in random methods[50][51], weight allocation can be achieved in several ways with different results. Two approaches were tested in this study. The first involves creating weights sequentially by generating random numbers ranging from 0.01, the minimum asset weight, to the maximum possible value that allows the remaining weights to be at least 0.01. Thus, each weight depends on the previously allocated weights.

The second approach, used in [12], involves generating 10 random numbers, scaling them so that their sum is 0.9 (a free proportion as it allows each asset to have a weight of at least 0.01), and then adding 0.01 to each number. Here the weights are allocated independently of each other.

To evaluate these approaches, each was used to create 1,000 random portfolios, which are plotted next to the UEF in [Fig.1]. It can be observed that sequential weight allocation leads to greater diversity between portfolios. Although sequential allocation can generate a portfolio that is far from the UEF, its best portfolios can form a line that is significantly closer to the UEF than the line formed by the best portfolios generated using independent allocation. Therefore, sequential allocation was chosen as the approach in this paper.

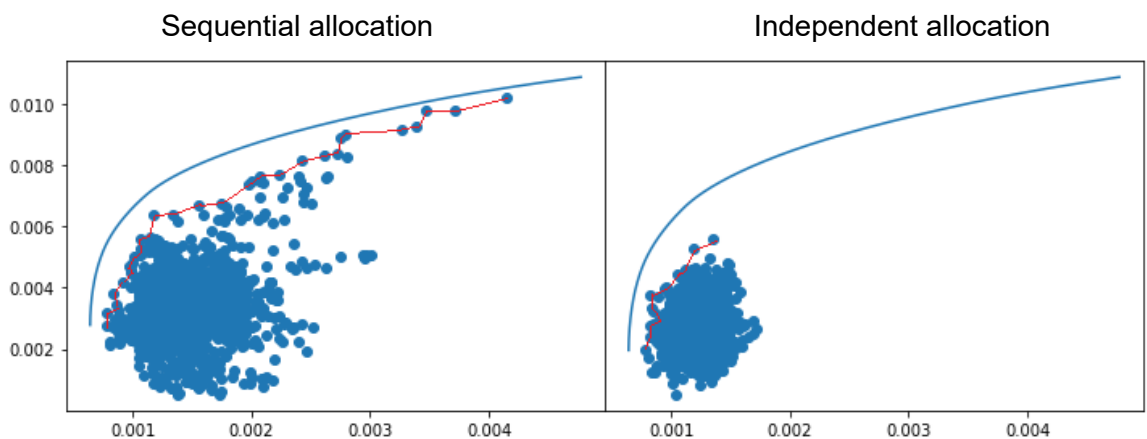

Fig.1: Comparing weight allocation approaches.

Sequential allocation          Independent allocation

# 5 Implementation of Metaheuristics

Three metaheuristics are implemented in this study: SA, TS, and GA. These are some of the most popular metaheuristics[10], and are therefore suitable candidates for testing the novel time-limited approach. Furthermore, the availability of datasets[39] previously used to test the same algorithms (although not in a time-limited context) provides a suitable benchmark for evaluating the performance of the metaheuristics.

## 5.1 Simulated Annealing

SA[C.2] starts with a single solution and iteratively selects random neighbouring solutions which, if accepted, replace the current solution. Its use for solving optimization problems originates in [52][53]. The algorithm is inspired by metallurgy, where annealing is the process of heating metal to a high temperature and then gradually cooling it down[54]. Similarly, SA algorithm incudes a temperature variable that decreases as the algorithm traverses through the search space. This temperature is proportional to the probability of the algorithm accepting a solution that is worse than a current solution, which enables SA to escape a local optimum[55]. An improved solution is always be accepted by the algorithm, while a solution that degrades the objective function by $\Delta$ at temperature $T$ is accepted with a probability of $e^{-\Delta/T}$.

The hyperparameters responsible for the performance of SA are known as the cooling schedule[17]. When tuning the initial temperature, $Tmax$, a balance must be found between too low a temperature at which the algorithm behaves like hill-climbing, only accepting better solutions, and too high a temperature at which SA accepts all solutions and therefore does not make progress. [17] recommends setting the initial temperature at a level at which the algorithm will accept 40-50% of all solutions. In this study, such an acceptance rate is achieved at a temperature of approximately 0.000015. However, tests have shown that a slightly higher initial temperature of 0.00005 resulting in the acceptance rate of 70%, leads to better results.

The cooling rate is determined by the constant $\alpha\ (0 < \alpha < 1)$, by which the temperature is multiplied after each iteration. [17] suggests using $\alpha$ of 0.5-0.99. In the context of constrained portfolio optimisation, $\alpha = 0.95$ is used in [12][13]. The higher the $\alpha$, the longer it takes for SA to reach lower temperatures. Slower cooling rates lead to better solutions, but require more iterations and therefore increase computation time[17]. Therefore, in this study, a higher value of $\alpha$ is used when running the algorithm over longer time periods. Specifically, $\alpha = 0.7$ is used for 1-second runs, $\alpha = 0.95$ is used for 5-second runs, and $\alpha = 0.995$ is used for 25-second runs.

Finally, most implementations of SA require a stopping condition. The most popular approach includes another hyperparameter, $Tmin$, which is the temperature at which the algorithm stops[17]. An alternative is to specify the number of iterations over which SA is executed, which is the approach taken in [12]. In this paper, the algorithm terminates after a specified amount of time. This is achieved using the $timeout$ hyperparameter, which specifies the number of seconds SA will run.

## 5.2 Tabu Search

TS[C.3] was proposed in [56], and involves selecting the best neighbouring solution that is not forbidden by the tabu list of recent solutions. Although it is intended to be a deterministic algorithm that evaluates all neighbouring solutions[17], such an approach would not be possible in the context of neighbourhood as defined in this study. Given that the weight of each asset is represented as a real number, the size of this neighbourhood dimension is infinite.

There are several ways in which TS can operate in such a neighbourhood. One approach is to select the first improving solution[12]. Another approach is to select the best solution from a subset of neighbours, consisting of some predetermined number of randomly chosen neighbouring solutions[57]. Both approaches are valid, but since the original implementation of the algorithm involves evaluating neighbouring solutions and selecting the best one, this study uses the approach that evaluates a subset of neighbours, since this best approximates the original approach.

Given that at each iteration TS generates and evaluates a new subset of neighbours, the size of this subset affects the performance of the algorithm. A larger subset increases the chance of finding a good solution, but is computationally demanding. On the other hand, a smaller subset that requires less time allows more iterations. A balance needs to be found between more iterations creating smaller subsets and fewer iterations creating larger subsets. Therefore, the size of the subset is one of the hyperparameters that affect the performance of TS. In this study, different sizes were tested at various stopping times. The best results were achieved when the size of the subset was proportional to the running time of the algorithm. At a computation time of 1 second, a subset of 10 solutions is used; at 5 seconds, a subset of 50 solutions; and at 25 seconds, a subset of 250 solutions.

The purpose of the tabu list is to prevent looping by forbidding the search to return to recently visited solutions[25]. Moreover, this list may allow TS to escape a local optimum[55]. It is usually implemented by maintaining a list of recent moves or attributes of moves. This is more efficient than storing complete solutions, which requires more space, as each solution will contain all assets, even if they have not been changed in the respective move. The tabu list functions as a buffer[22], because its length is fixed and at each iteration the most recent move is added and the oldest move in the list is removed. The length of the tabu list is known as the tabu tenure[17]. In this paper, tabu tenure values 3, 5 and 7 were tested[B.1]. A value of 3 resulted in the lowest MPE and was therefore chosen.

Given that in this study a neighbourhood is defined in two dimensions, separate tabu lists can be allocated for both definitions. Moreover, in each dimension the asset can be changed in two ways - in one dimension the asset can be added to or removed from the portfolio and in the other dimension its weight can be increased or decreased. This means that there are four possible tabu lists: a list of assets recently placed in the portfolio; a list of recently removed assets; a list of assets which weight has been reduced; and a list of assets which weight has been increased. Note that when changing the weights of assets already in the portfolio, the exact change in proportion is not considered, and the only information entered into the list is whether the weight has increased or decreased. This is because the weight is represented by a real number and can therefore be changed in an infinite number of ways.

In this study, all 16 combinations of the four tabu lists were tested at various time limits[B.2]. The best results are achieved with a single tabu list containing assets which weight has been reduced and thus cannot be increased during the tabu tenure. This is the tabu list used in this study.

## 5.3  Genetic Algorithm

GA[C.4], first developed in [58], is an evolutionary algorithm inspired by natural selection and sexual reproduction[17]. It first generates a set of random solutions, known as a population. At each iteration it selects a pair of solutions based on the objective function – the smaller the objective function, the higher the probability of being selected. These solution pairs are combined to produce "offspring" solutions in a procedure known as crossover. The offspring inherit the characteristics of both "parent" solutions[55]. Moreover, every "child" has a chance to undergo a mutation that changes certain aspects of the solution. Finally, a replacement scheme dictates which offspring enter the population, and which parent solutions leave it[17]. The metaheuristic iterates until the stopping criterion is reached. In this study, the algorithm stops after a certain time limit.

One of the hyperparameters that require tuning in this metaheuristic is the population size. While a larger population may lead to better solutions, it also increases time complexity[17]. A population size of 20-100 is recommended in [17], while [59] considers a population of about 50-200 to be suitable. In the context of constrained portfolio optimisation in [12][13], a population size of 100 is used. However, given that these two papers do not restrict the running time of metaheuristics, their exact approach will not be used. In this study, different population sizes were tested at different stopping times, and it was found that while smaller populations perform better at shorter time periods, larger populations perform better at longer time periods. Specifically, a population of 20 portfolios is used for 1-second runs, a population of 50 for 5-second runs, and a population of 200 for 25-second runs.

There are several approaches to creating the initial population. In [12], 100 random portfolios are generated to form a population of 100 members. This study uses a novel approach in which 1,000 random portfolios are generated and an initial population of 20-200 members (depending on the time limit) is created from the portfolios with the smallest objective function. When compared to the approach in [12], the novel approach leads to a significant improvement in solution quality.

In this paper, as in [12], the parent solutions are chosen using binary tournament selection. This involves selecting four random solutions, dividing them into two sets of two, and choosing from each set the solution with the smallest objective function. These two solutions are then subjected to uniform crossover to produce offspring. An asset present in both parents is present in the offspring with the corresponding weight randomly taken from one parent, while an asset present in one parent (and its corresponding weight) is present in the offspring with a 50% chance. Given that weights are taken from both portfolios, it is unlikely that their sum will equal 1 in the offspring portfolio, and so the weights will require scaling. This is achieved by multiplying the fraction of each weight above 0.01 (the minimum weight constraint) by the number needed for the sum of the weights to equal 1.

Each offspring may then be subject to mutation, a process that alters certain aspects of the solution. Although [17] recommends applying mutation with a low probability of 0.001-0.01, in [12] every offspring portfolio undergoes a mutation in which the weight of a randomly selected asset is increased or decreased by 10%. A different approach is taken in [13], where an asset in the portfolio is replaced by an asset not in the portfolio with a probability of 0.03. In this study, the neighbourhood is two-dimensional, and so the mutation operator implemented here can change both asset selection and weight distribution with a probability of 0.1 each.

Given that the population size does not change during permutations, each offspring is placed in the population in place of the portfolio with the worst objective function. This is known as a steady-state population replacement strategy[17], and it is the approach used in [12] and in this paper.

## 6    Computational Results

### 6.1    Programming Language

Although [25] emphasises the importance of choosing the suitable programming language to implement metaheuristics, it may be noted that in comparative studies such as this one, the absolute computation time is not as important as the performance of algorithms in relation to each other. The aim of this study is to find out which algorithm offers the best performance in a short computation time, not to establish a performance benchmark. This study uses a personal computer to test the metaheuristics, and achieving better performance on more powerful hardware is not difficult. Nevertheless, given that this paper tests the short-term performance of metaheuristics under specific time limits, the algorithms must be implemented efficiently to achieve meaningful results.

Python was chosen as the programming language. It is a dynamically typed language[60], i.e. type checking is done at runtime rather than at compile time, and therefore a program written in Python is slower to execute than an identical program written in a statically typed language like Java[61]. NumPy, the Python library for numerical analysis[62], partially alleviates this problem by providing efficient containers and operators for fast mathematical operations[63]. NumPy's ability to perform mathematical operations on arrays without using loops is used in this paper to implement scaling of offspring weights in GA.

Care has been taken to ensure that suitable data structures are used. For example, in GA the offspring is placed in the population in place of the individual with the worst objective function. Therefore, the population must be implemented as a container that maintains a sorted order. Using a list would result in a linear $O(n)$ time for inserting an offspring into a sorted population[64]. A better alternative is to use a heap structure where the insertion of an offspring is done in logarithmic $O(log\ n)$ time[65]. In this study, the heap is implemented using heapq module[66].

## 6.2 Solution Quality Measurement

The quality of the solution portfolios is assessed by measuring deviation from the UEF. There are multiple ways to do this. One approach is to measure the Euclidean distance from the portfolio to the nearest point in the UEF. In [12], the percentage deviation of the portfolio from the UEF along both the x- and y-axes is measured, and the smallest measurement is taken. Since the UEF is not continuous, linear interpolation is used when there is no point on the UEF with an identical $x$ or $y$ dimension. For linear interpolation to work, the UEF must contain points with higher or equal, and lower or equal values than the solution portfolio in both $x$ and $y$ dimensions. If this is not the case, no value is calculated in that dimension, and the portfolio is evaluated on the other dimension if it is available.

The problem with this approach is that it disproportionately penalises portfolios that are outside the UEF range. As can be seen in [Fig.2], the deviation of portfolio A would be measured along its shorter x-axis distance from the UEF, whilst the deviation of Portfolio B, which is offering a lower return but also a lower risk, would be measured on its longer y-axis distance from the UEF, as there are no points in the UEF with a lower return. Thus, B is calculated as having a significantly larger error than A, even though the portfolios are relatively similar. This issue is particularly important when evaluating portfolios at the lower left end of the frontier, as solution portfolios often offer returns that are slightly lower than that of the UEF portfolio with the lowest return. Indeed, in [12] several portfolios are below the left edge of the frontier[Fig.2].

Fig.2: Portfolios on the edge of the UEF range, adapted from [12].

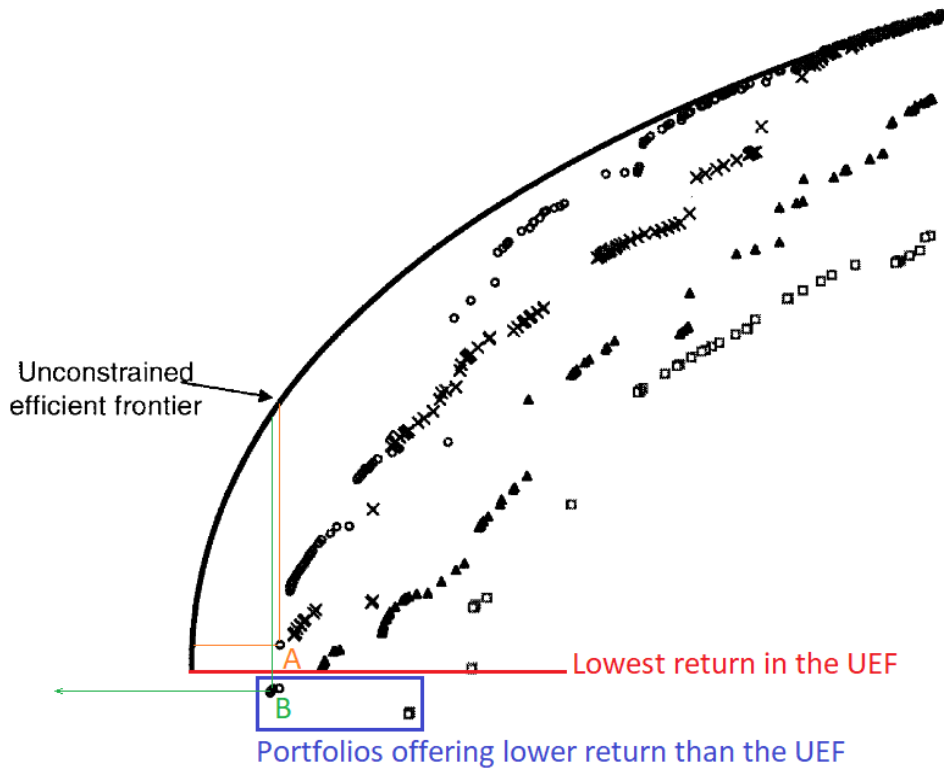

This study proposes a solution to this problem by combining the linear interpolation approach used in [12] with the Euclidean distance approach: if the UEF does not contain a point with a lower return, calculate the error by taking the shortest of the straight-line distance along the y-axis and the Euclidean distance to the point on the UEF with the lowest return. To test this approach, four hypothetical portfolios offering the same level of risk (at arbitrarily chosen standard deviation of 0.027) and slightly different levels of return were evaluated, with each portfolio returning $1 \times 10^{-8}$ less than the other[Tab.4]. The level of return of two portfolios is greater than or equal to the lowest return in the UEF, while the other two portfolios offer lower returns.

Tab.4: Comparing evaluation approaches.

| Mean return | Percentage error | |
|---|---|---|
| | Linear interpolation approach | Combined approach |
| 0.00278435 | 6.53916112765539 | 6.53916112765539 |
| 0.00278434 (lowest in the UEF) | 6.539161209732704 | 6.539161209732704 |
| 0.00278433 | 43.856524118314816 | 6.5391612401485695 |
| 0.00278432 | 43.856725759197474 | 6.539161240417628 |

As can be seen in [Tab.4], the error measurement for the two portfolios that are outside the UEF range increases dramatically when evaluated using the linear interpolation approach proposed in [12]. We believe that it is not the intended behaviour, given that the difference in returns is minute and the risk is the same across the four portfolios. When evaluating portfolios using the combined approach proposed in this paper, the percentage error remains very similar, as would be expected for portfolios with identical risk and marginally different returns.

In addition, it should also be noted that although the UEF presented in the OR-Library dataset contains the variance of the return[43], [12] uses the standard deviation when calculating the percentage deviation, because the standard deviation is expressed in the same units as the original data[67]. Using variance, which is the square of the standard deviation, will not result in commensurate units. The Euclidean calculation used in this paper requires that risk and return are measured in the same units, and therefore this study also uses standard deviation instead of variance.

## 6.3 Results

All metaheuristics tested in this study are written in Python 3.10.7 64-bit and run on a 16GB RAM, Intel Core i7 2.60GHz CPU personal computer running the Windows 11 Home 64-bit operating system. Given that metaheuristics operate stochastically, their results are not consistent[22]. To account for this, each algorithm was tested several times - 50 times on 1 and 5-second tests and 30 times on 25-second tests. The resulting MPEs are added together, and the average value is taken.

Note that [12] evaluates solutions in two ways. First, the percentage error is measured for the portfolios associated with each given $\lambda$ (since there are 50 values of $\lambda$, 50 portfolios are evaluated). Secondly, the dominated portfolios are removed, and MPE is only measured for the best portfolios. A dominated portfolio is a portfolio that offers both lower return and higher risk than another portfolio in the solution set[68]. In [13] the portfolios are evaluated in the second way only. In this paper the solutions are evaluated in the first way, where every portfolio in a solution set is assessed, leading to results that are less affected by chance. The second mode of assessment allows for a hypothetical situation where a solution containing one good portfolio dominating many poor portfolios is assessed on the quality of a single portfolio. Given that all the metaheuristics implemented in this study operate stochastically, there is always the possibility that an algorithm will find an unusually good portfolio, and such an event should not greatly affect the algorithm's overall performance score.

The results in [Tab.5] show MPEs calculated for the solution set of 50 portfolios (one per $\lambda$) found by each metaheuristic on each dataset in 1, 5 and 25 seconds. Note that the time limit is set for the entire solution set of 50 portfolios, resulting in 20ms, 100ms and 500ms for each individual portfolio. The results also include the average MPE achieved by time-limited metaheuristics across all five datasets. This makes it easier to compare algorithms and is common practice in related studies[12][13][30]. The lowest MPE values achieved for each dataset are shown in bold.

Tab.5: Test problem results (MPEs).

| Index | Number of assets | 1 second | | | 5 seconds | | | 25 seconds | | |
|---|---|---|---|---|---|---|---|---|---|---|
| | | SA | TS | GA | SA | TS | GA | SA | TS | GA |
| Hang-Seng | 31 | 1.5919 | 1.8770 | 2.7368 | 1.1219 | 1.2043 | 1.3624 | **1.0950** | 1.1161 | 1.1067 |
| DAX-100 | 85 | 5.7169 | 5.8466 | 7.2523 | 2.4445 | 2.6614 | 3.7058 | **2.3280** | 2.3989 | 2.4741 |
| FTSE-100 | 89 | 2.1363 | 2.7262 | 4.3982 | 0.8898 | 1.0920 | 2.0266 | **0.8456** | 0.9024 | 0.9739 |
| S&P-100 | 98 | 4.0871 | 4.6423 | 5.9036 | 1.5256 | 1.8607 | 2.7643 | **1.3869** | 1.4619 | 1.5392 |
| Nikkei-225 | 225 | 7.1977 | 8.4752 | 9.4205 | 1.1911 | 2.1680 | 3.9640 | **0.5980** | 0.8226 | 1.1505 |
| Average | | 4.1460 | 4.7135 | 5.9423 | 1.4346 | 1.7973 | 2.7646 | 1.2507 | 1.3404 | 1.4489 |

# 7 Analysis

## 7.1 Discussion

The relative performance of the algorithms remains almost unchanged regardless of the time limit: two S-metaheuristics outperform GA, with SA achieving the smallest MPEs. The only exception is the Hang-Seng dataset with a time limit of 25 seconds, where GA achieves a lower MPE than TS. This confirms the hypothesis that S-metaheuristics provide higher quality solutions than P-metaheuristics for the same computation time.

Comparing the results obtained for different time limits, it can be concluded that metaheuristics are not capable of finding a competitive solution in one second. However, in 5 seconds SA is able to find a solution that is within 10% of the best MPE obtained for four datasets with less than 100 assets. In 5 seconds, TS finds a solution that is within 10% of the best MPE obtained only for the Hang-Seng dataset with the lowest number of assets. GA is unable to find a competitive solution in 5 seconds.

The progress made by SA in different time limits is visualised in [Fig.3]. It can be seen that longer time limits allow the algorithm to find portfolios closely tracing the UEF.

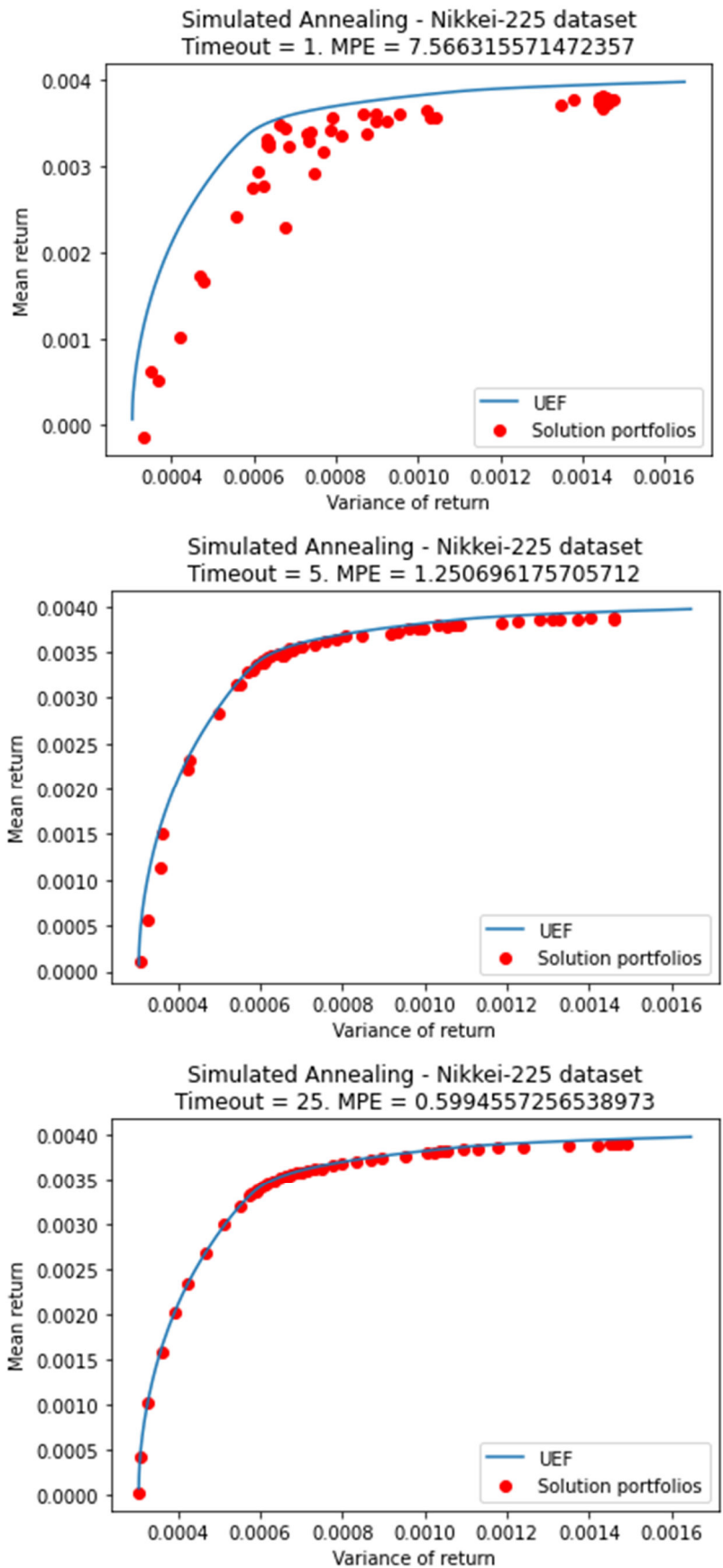

Fig.3: Visualisation of SA portfolios.

The performance of metaheuristics can also be assessed by measuring the percentage improvement (relative change) achieved from one time limit to the next[Tab.6]. Although all metaheuristics achieve better results in longer time limits, the rate of improvement usually decreases with time. Thus, a conclusion can be reached that as the algorithm approaches the optimal solution, the rate of improvement decreases. Indeed, this behaviour follows the shape of the convergence curve common to all metaheuristics[69].

Tab.6: Percentage improvement.

| Index | Number of assets | Metaheuristic | 1-5 sec | 5-25 sec |
|---|---|---|---|---|
| Hang-Seng | 31 | SA | 29.52% | 2.40% |
| | | TS | 35.84% | 7.32% |
| | | GA | 50.22% | 18.77% |
| DAX-100 | 85 | SA | 57.24% | 4.77% |
| | | TS | 54.48% | 9.86% |
| | | GA | 48.90% | 33.24% |
| FTSE-100 | 89 | SA | 58.35% | 4.97% |
| | | TS | 59.94% | 17.36% |
| | | GA | 53.92% | 51.94% |
| S&P-100 | 98 | SA | 62.67% | 9.09% |
| | | TS | 59.92% | 21.43% |
| | | GA | 53.18% | 44.32% |
| Nikkei-225 | 225 | SA | 83.45% | 49.79% |
| | | TS | 74.42% | 62.06% |
| | | GA | 57.92% | 70.98% |

The speed at which the rate decreases is influenced by the number of assets in the stock index and the choice of a metaheuristic. Finding a viable portfolio in a stock index with more assets requires more computation time, as seen in comparable studies[12][13][32]. SA offers the lowest percentage of improvement between 5-25 seconds for all datasets. Given that SA also achieves the smallest MPEs it can be concluded that SA converges faster than other metaheuristics, and thus provides the best short-term performance. Likewise, the inferior performance of GA can be understood in the context of GA having the highest percentage of improvement between 5-25 seconds for all data sets. It can be assumed that at the 25-second limit GA is still making significant progress, and a higher quality solution will be achieved at some later point.

When comparing with the results from [12] in [Tab.1], it can be noted that MPEs obtained under the 25-second time limits in this study are smaller in all but one case, the GA on Nikkei-225. Moreover, by comparing the average MPE scores, it can be seen that S-metaheuristics implemented in this study can outperform the results obtained in [12] in 5 seconds.

Compared to the results obtained in [13], only the SA implemented in this study shows better results: in [13] the average MPE is 1.4391 compared to 1.2507 in this study. This can be explained by the effectiveness of the subset optimisation step in [13], which is particularly powerful when dealing with a large number of solutions – a population in GA or a neighbourhood subset in TS. However, the running time of metaheuristics in [13] is significantly longer: on the Nikkei-225 dataset, 604 seconds were required for SA.

## 7.2 Limitations

One potential risk to the validity of the results of this study is that the implementation of the three metaheuristics may be of different quality. This could result in an algorithm outperforming another due to the way it is coded rather than because it is better suited for portfolio optimisation. To reduce this risk, this study adopts code reuse, which is a principle of object-oriented programming[70]. If several algorithms have common components with the same functionality, they are implemented through function calls to the same methods. For example, GA and SA both use replace_asset and change_weights, while GA and TS use find_starting_portfolios to find the initial solution for each $\lambda$.

The fact that a single metaheuristic can be designed in a myriad of ways presents another limitation, as different versions of an algorithm can achieve contrasting results. For example, features that can be added to TS include medium- and long-term memory, responsible for intensification and diversification respectively[17]. Moreover, each of these features can be implemented in different ways. Although numerous design decisions were evaluated in this study, it is not possible to test all existing approaches. Therefore, testing other versions of the algorithms is an opportunity for future research.

Another limitation of this study is in the risk that the implemented metaheuristics are not fast enough for real-world use cases. Given that the execution time for market orders is less than a second[6], a solution found in 5 seconds may already be too late. However, it is important to note that in this study the solution consists of 50 portfolios, so a solution found in 5 seconds results in one portfolio found in 100ms. Furthermore, the computation time of all metaheuristics can be improved without changing the algorithms by using more powerful hardware and by implementing the algorithms in a faster compiled language[71]. Indeed, in automated trading, where speed is critical, C++ is usually used[72]. It is therefore possible that the algorithms tested in this study could be useful in algorithmic trading if implemented on a suitable system.

## 7.3 Conclusion and Future Work

Given that past studies of metaheuristics in the context of constrained portfolio optimisation do not keep computation times constant[11]–[14], the comparative performance of time-limited metaheuristics has been identified as a viable research gap. Also noting the relative underutilisation of S-metaheuristics[7][10], this study identified two S-metaheuristics (SA and TS) and one P-metaheuristic (GA) and examined their performance under time limits on a dataset previously used in comparable studies[12][13]. SA was found to be the algorithm that performs best under time constraints.

In addition, several novel approaches to implementing metaheuristics have been evaluated in this study. The introduction of two-dimensional neighbourhoods in S-metaheuristics has led to the investigation of multiple tabu lists. A novel approach to creating the initial population was tested in GA. As a result, all metaheuristics achieved improved MPEs compared to [12]. Furthermore, the relationship between hyperparameters and time limits was explored, leading to the establishment of recommended parameter values for specific time limits.

Future research could develop the time-limited approach by testing other metaheuristics and experimenting with different implementations of the algorithms evaluated in this study. For example, quadratic subset optimisation implemented in [13] significantly improves the performance of P-metaheuristics, and thus can show competitive results under time constraints. Another way of extending the research in this paper would be to use optimum hardware and a compiled programming language to establish performance benchmarks.

# 8 Ethics

The dataset used in this study is publicly available market data obtained from the OR-Library, openly accessible for use under the MIT licence[40]. Therefore, the ethical criteria of data confidentiality and informed consent do not apply. The study does not use data related to any biological organisms and has no known military applications. Furthermore, the research is not funded by, or conducted in collaboration with, an organisation that may bring the University of York into disrepute. However, as trading in shares may be considered unethical[73], various ethical concerns need to be considered.

Ethics is concerned with what is perceived as right, and so unethical things can be permitted by law[74]. For example, the tobacco and alcohol industries are known to encourage harmful behaviour. The companies in these sectors are supported through the sale of shares, referred to as "sin stocks"[75]. The market indices used in this study are known to contain sin stocks. For example, British American Tobacco is a component of the FTSE-100[76] and arms manufacturer Lockheed Martin is in the S&P-100[77]. When optimising a portfolio, sin stocks are likely to be chosen because they often have a higher expected return than comparable shares[78].

However, the focus of this study is the performance of metaheuristics, not the creation of a real portfolio. Thus, the use of a well-known, open-source dataset is appropriate for the task. Furthermore, since the identity of individual stocks is hidden in the dataset used[12], it is not known whether sin stocks are present in the selected portfolios. However, future studies can address this issue by using a dataset that meets some ethical criteria. There are market indices specifically designed for ethical investing, such as FTSE4Good, featuring businesses demonstrating "Environmental, Social and Governance practices"[79].

Another issue to consider is accountability. All investments carry the risk of losing money[80]. Furthermore, algorithms can behave unexpectedly and make bad decisions, despite the best intentions of the authors. Building a real investment portfolio using metaheuristics can lead to financial losses. However, this study does not aim to develop a program appropriate for automated trading. The decision to invest is left to the investor if metaheuristics are used in an advisory role. A comparison can be drawn with medical expert systems, which are regarded by law as being analogous to textbooks, and where doctors must use their own discretion in determining whether to follow the suggestions offered by the system[55].

# Appendix A: Artefacts

The zipped folder "Artefacts.zip" contains the following files and folders that serve as artefacts:

1) Subfolder "data" containing two subfolders:
   a) Subfolder "stocks" containing 5 files: "port1", ..., "port5".

      The format of these data files is:
      number of assets (N)
      for each asset i (i=1,...,N):
          mean return, standard deviation of return
      for all possible pairs of assets:
          i, j, correlation between asset i and asset j

   b) Subfolder "UEFs" containing 5 files: "portef1", ..., "portef5".

      The format of these files is:
      for each of the calculated points on the unconstrained frontier:
          mean return, variance of return

2) "IRP.ipynb" – Jupyter Notebook containing the code required to reproduce this study.
3) "Results.txt" – containing unrounded MPEs of all tests.
4) "Alexander_Nikiporenko_Self_Assessment_Form.pdf" – ethics form, signed by the supervisor.

# Appendix B: Hyperparameter Tuning

Tab.B.1: Tabu tenure.

| Tabu tenure | MPE | | | |
|---|---|---|---|---|
| | 1 sec | 5 sec | 25 sec | Average |
| 3 | 4.7980 | 1.7717 | 1.3579 | 2.7319 |
| 5 | 4.8140 | 1.8224 | 1.3688 | 3.2513 |
| 7 | 4.7847 | 1.8911 | 1.3576 | 3.7584 |

Tab.B.2: Tabu lists.

| Tabu lists | | | | MPE | | | |
|---|---|---|---|---|---|---|---|
| Assets in | Assets out | Weight up | Weight down | 1 sec | 5 sec | 25 sec | Average |
| - | - | - | - | 4.7120 | 1.8221 | 1.3460 | 2.6267 |
| - | - | - | ✓ | 4.6320 | 1.7959 | 1.3402 | 2.5894 |
| - | - | ✓ | - | 4.7505 | 1.8583 | 1.3440 | 2.6509 |
| - | ✓ | - | - | 4.7707 | 1.8241 | 1.3922 | 2.6623 |
| ✓ | - | - | - | 5.3829 | 2.1154 | 1.4984 | 2.9989 |
| - | - | ✓ | ✓ | 4.8129 | 1.8383 | 1.3574 | 2.6695 |
| - | ✓ | - | ✓ | 4.8162 | 1.8391 | 1.4195 | 2.6916 |
| ✓ | - | - | ✓ | 5.2579 | 2.2244 | 1.5178 | 3.0000 |
| - | ✓ | ✓ | - | 5.0384 | 1.8425 | 1.3871 | 2.7560 |
| ✓ | - | ✓ | - | 5.4220 | 2.2126 | 1.4843 | 3.0396 |
| ✓ | ✓ | - | - | 5.3171 | 2.2478 | 1.6957 | 3.0869 |
| - | ✓ | ✓ | ✓ | 4.7590 | 1.8746 | 1.4173 | 2.6836 |
| ✓ | - | ✓ | ✓ | 5.3270 | 2.1771 | 1.5206 | 3.0082 |
| ✓ | ✓ | - | ✓ | 5.4072 | 2.2617 | 1.6393 | 3.1027 |
| ✓ | ✓ | ✓ | - | 5.3519 | 2.2518 | 1.6597 | 3.0878 |
| ✓ | ✓ | ✓ | ✓ | 5.4326 | 2.2641 | 1.6613 | 3.1193 |

# Appendix C: Pseudocode

## Algorithm C.1: Finding Starting Portfolios

```
FIND_STARTING_PORTFOLIOS(n)
1      Input: Rndm = 1000 random portfolios
2             Lmbd = n lambda values evenly distributed between 0 and 1
3      P = array [1 .. n]
4      for l in Lmbd:
5              f_values = array [1 .. 1000]
6              for r in Rndm:
                       /* Objective function is measured for every random portfolio */
7                      f_values.APPEND( EVALUATE(r) )
               /* Portfolio with a lowest objective function is added to the P array */
8              P.APPEND( MIN(f_values) )
9      return P
```

## Algorithm C.2: Simulated Annealing

```
SIMULATED_ANNEALING
1      Input: Prtf = FIND_STARTING_PORTFOLIOS(50)
2             Lmbd = 50 lambda values evenly distributed between 0 and 1
3             α = value depending on time limit
       /* temperature */
4      T = Tmax
       /* replace boolean is responsible for switching between neighbourhoods */
5      replace = True
6      while time < (time_start + timeout):
7              for l in Lmbd:
8                      p = Prtf[l]
9                      if replace:
10                             p_candidate = p.REPLACE_ASSET()
11                     else:
12                             p_candidate = p.CHANGE_WEIGHTS()
13                     ΔE = EVALUATE(p_candidate) - EVALUATE(p)
                       /* if the candidate solution is better, it is accepted */
14                     if ΔE <= 0:
15                             p = p_candidate
                       /* otherwise it is accepted with a probability dependent on T */
16                     else:
17                             p = p_candidate with a probability e^(-ΔE/T)
18             T = T * α
19             replace = not replace
20     return Prtf
```

## Algorithm C.3: Tabu Search

```
TABU_SEARCH
1       Input: Prtf = FIND_STARTING_PORTFOLIOS(50)
2              Lmbd = 50 lambda values evenly distributed between 0 and 1
3              subset = value depending on time limit
        /* Tabu list, containing attributes of the solutions */
4       Tabu = deque [ ]
        /* replace boolean is responsible for switching between neighbourhoods */
5       replace = True
6       while time < (time_start + timout):
7               for l in Lmbd:
8                       p = Prtf[l]
9                       options = array [ ]
                        /* populating options array with candidate solutions */
10                      if replace:
11                              repeat while LENGTH(options) < subset:
12                                      options.APPEND( p.REPLACE_ASSET() )
13                      else:
14                              repeat while LENGTH(options) < subset:
                                        /* only a solution that is not in Tabu list
                                           can become a candidate */
15                                      repeat while p_candidate not in Tabu:
16                                              p_candidate = p.CHANGE_WEIGHTS()
17                                      options.APPEND(p_candidate)
                        /* a candidate with a lowest objective function is accepted */
18                      p = MIN( EVALUATE(options) )
                        /* attributes of the accepted solution are added to the Tabu list */
19                      Tabu.APPEND_RIGHT(p)
                        /* the length of the Tabu list is kept constant */
20                      if LENGTH(Tabu) > 3:
21                              Tabu.POP_LEFT()
22              replace = not replace
23      return Prtf
```

## Algorithm C.4: Genetic Algorithm

```
GENETIC_ALGORITHM
1       Input:  Rndm = 1000 random portfolios
2               Lmbd = 50 lambda values evenly distributed between 0 and 1
3               pop_size = value depending on time limit
        /* 50 populations are maintained, 1 for each lambda */
4       populations = [1 .. LENGTH(Lmbd) ]
5       for l in Lmbd:
                /* Objective function is measured for every random portfolio */
                /* Random portfolios are sorted according to their objective function */
6               evaluated = EVALUATE(Rndm).SORT()
                /* pop_size best random portfolios become the population */
                /* each population is a heap */
7               populations[l] = HEAPIFY( evaluated[1 .. pop_size] )
8       while time < (time_start + timout):
9               for l in Lmbd:
                        /* Parents are selected using binary tournament */
10                      parents = 4 random portfolios from populations[l]
11                      p1 = MIN( EVALUATE( parents[1, 2] ) )
12                      p2 = MIN( EVALUATE( parents[3, 4] ) )
                        /* A child is created */
13                      child = CROSSOVER( p1, p2 )
14                      child.MUTATE()
                        /* 10% chance to REPLACE_ASSET() */
                        /* 10% chance to CHANGE_WEIGHTS() */
15                      populations[l].HEAP_PUSH(child)
                        /* Child is added to the population */
16                      populations[l].HEAP_POP()
                        /* A population member with the
                        worst objective function is removed */
17      return MIN( EVALUATE(populations) )
```

# Appendix D: Glossary

| Term | Definition |
| --- | --- |
| ACO | Ant colony optimisation |
| GA | Genetic algorithm |
| ILS | Iterated local search |
| MPE | Mean percentage error |
| P-metaheuristic | Population-based metaheuristic |
| PSO | Particle sward optimisation |
| SA | Simulated annealing |
| S-metaheuristic | Single-solution metaheuristic |
| TS | Tabu search |
| UEF | Unconstrained efficient frontier |
| VNS | Variable neighbourhood search |
| 2PLS | Two-phase local search |